# Machine and quantum learning for diamond-based quantum applications

Dylan Geoffrey Stone[1] and Carlo Bradac[1,*]

[1] Trent University, Department of Physics & Astronomy, 1600 West Bank Drive, Peterborough, ON, K9L 0G2.
* corresponding author: Carlo Bradac, carlobradac@trentu.ca

**Abstract**
In recent years, machine and quantum learning have gained considerable momentum sustained by growth in computational power and data availability and have shown exceptional aptness for solving recognition- and classification-type problems, as well as problems that require complex, strategic planning. In this work, we discuss and analyze the role machine and quantum learning are playing in the development of diamond-based quantum technologies. This matters as diamond and its optically-addressable spin defects are becoming prime hardware candidates for solid state-based applications in quantum information, computing and metrology. Through a selected number of demonstrations, we show that machine and quantum learning are leading to both practical and fundamental improvements in measurement speed and accuracy. This is crucial for quantum applications, especially for those where coherence time and signal-to-noise ratio are scarce resources. We summarize some of the most prominent machine and quantum learning approaches that have been conducive to the presented advances and discuss their potential for proposed and future quantum applications.

## 1. Introduction

In the last few years, machine learning (ML) has undergone tremendous progress, mainly driven by increased computational power and access to large amounts of data. Some of its most impressive results include solving complex problems such as image classification,[1] face[2] and speech recognition,[3] language processing[4] and strategic gaming,[5–7] as well as advancing discovery in fields ranging from material engineering[8–10] to computational chemistry,[11,12] photonics[13–15] and microscopy.[16] Machine learning has additionally been used in the design and optimization of semiconductor quantum devices where its proclivity for fast and efficient automation has made it ideal for large-scale fabrication, characterization and fine-tuning operations.[17–19] In conjunction with ML, the parallel field of quantum machine learning (QML) is also rapidly developing, fueled by the idea that quantum-based computational paradigms could both speed up and improve the ability to find statistical patterns between sets of data, which is at the core of machine learning.[20–22]

In this review, we specifically look at how developments in machine and quantum learning are advancing the field of diamond-based quantum sensing and metrology. Color centers in diamond have quickly become archetypal systems for quantum technologies owing to their optically-addressable spins, long coherence times, room-temperature operation and the ability to create long-range entanglement through photons.[23] Notably, as some of the proposed diamond-based applications have become established, the field is organically shifting its focus from carrying out proof-of-principle experiments to engineering practical realizations that are technology-oriented. This is fostering two parallel and complementary streams of research. The first is hardware-driven: it relies on progress in material science and aims at fabricating better diamond material with tailored properties such as, e.g., controlled density of color centers, and access to individual emitters with high photon-extraction efficiencies and desirable spectral/coherence properties.[24,25] The second is software-driven: it revolves around designing novel experimental protocols for data acquisition and data analysis to maximize, e.g., speed and resolution of specific measurements.[26,27] This second focus is where machine and quantum learning find their natural fit, for they offer novel computing approaches that—even in their simplest realizations—can achieve measurement speeds and/or accuracies ~1-2 orders of magnitude higher than those of traditional statistical approaches. Interestingly however, ML is also indirectly becoming a powerful tool for the first focus concerning material science. Thanks to the distinctive efficiency in establishing correlations and finding patterns, models and simulations powered by ML can effectively guide the choice of optimal parameters for material and device engineering. In this review we will analyze all these aspects.

The paper is organized as follows. We first introduce (cf. § 2) some key background concepts about ML and QML that are relevant for the scope of this review. We then present a selection of demonstrations where ML and QML have produced significant advances in the development and improvement of diamond-based quantum applications (cf. § 3). We go beyond merely presenting the results and use these demonstrations as an opportunity to analyze—critically—how the proposed ML and QML algorithms work in each specific experiment. We highlight the merits and shortcomings of the various algorithms and discuss their potential application to unexplored aspects of quantum metrology. We then present a summary of some of the most prominent ML and QML methods that have been conducive to the advances we discussed (cf. § 4). We review these methods in a broader context, beyond their specific application to quantum technologies. This is functional to the goal of this work to be a reference for a mixed audience of experts in quantum science and machine learning, serving as a bridge between the two communities. We conclude the review by summarizing the potential of ML and MQL for proposed and future quantum applications (cf. § 5).

## 2. Background
### 2.1 Artificial intelligence and machine learning
In its broader scope, artificial intelligence (AI) deals with understanding and abstracting—ultimately with the goal of replicating—the distinctive features of human intelligence. These may range from specific activities such as perception, motion, navigation and language processing[28] to more general capabilities including planning and problem solving.[29] In essence, AI aims to create intelligent (artificial) agents, that can interact with an unknown environment through learning from, and reacting to it—in a capacity similar to that of a human. This requires the agent to display aptitude for knowledge accumulation and representation (learning), combined with logical reasoning and decision making (reacting).

Machine learning is a specialized subfield of artificial intelligence. It focuses on the technological development of algorithms that—formally—can learn from experience (E) with respect to some task (T) and some performance indicator (P), and such that the performance on T, as measured by P, improves with experience E.[30] This is the focus of this review. We discuss how ML/QML algorithms are improving the performance of certain tasks—e.g. monitoring an observable, classifying the state of a quantum system or tracking its dynamic evolution—in the context of diamond-based quantum applications. The performance of ML/QML algorithms is directly compared to that of traditional approaches based, e.g., on least-square data fitting and is quantified through metrics such as measurement speed, sensitivity, resolution and accuracy. Note that in general, these performance metrics fall under two categories: *sample* and *time complexity*.[31] Sample complexity is the smallest number of data points required to learn a function within a specified accuracy. Time complexity is the runtime of the learning algorithm—usually considered efficient if it is polynomial in the dimension of the elements and inverse polynomial in the error parameters. Within this framework, ML aims at designing statistical estimators (that are, ideally, computationally fast and require few observations) from known data, to fit or predict accurately unseen data. This is known in ML as the *generalization capability* of the learned model and it distinguishes ML—albeit conceptually rather than

fundamentally—from standard optimization data fitting, which instead focuses on determining the parameters of an inferred model from a series of observations (cf. § 3).[32]

*2.2 Supervised, unsupervised and reinforcement learning*

Generally speaking, machine learning deals with the realization of algorithms that can learn from data and make predictions about it. Given a set of input and output observations, ML algorithms are functions that map the input(s) to the output(s) within a specified level of accuracy—usually defined by statistical estimators such as variance or standard deviation from the mean. Machine learning approaches are often classified in three main categories: *supervised*, *unsupervised* and *reinforcement learning*,[33,34] briefly outlined below. Hybrid approaches involving different combinations of these categories also exist.[35,36]

Supervised learning. It is a learning-by-example strategy. The ML algorithm has access to a certain number of *labeled points,* so-called *training set(s)*, $\{(x_j, y_j)\}_i$, where $x_j$ are data points ($n$-dimensional vectors of descriptive attributes, often referred to as *features*) and $y_j$ are *labels* (the observable(s), e.g. binary or real values, to be forecast). For the training sets both features and labels are known: the ML algorithm uses this knowledge to infer a labeling rule $x_j \to y_j$ that predicts the label $y_j$ from an unknown set of features $x_j$, different from the training set. In this approach, it is common to reserve a second set of *labeled points*, the so-called *testing set*, to validate the labeling rule, as the features and labels of the testing set are also known. More formally, the ML algorithm infers the conditional probability $P(Y = y | X = x)$ that assigns labels to features based on a certain number of samples (training and testing sets) from the joint distribution $P(X, Y)$.

Unsupervised learning. In this case, the algorithm has access to the data points (features), without the labels. The goal is for the ML algorithm to identify underlying patterns or structures the features might display. Formally, this consists in inferring properties of the distribution $P(X = x)$ based on a certain number of samples, with respect to a rule defined by the user. An example of unsupervised learning is the assignment of data to a group or cluster. The process is unsupervised as the clusters and/or their numbers are obtained through an optimization process without prior labeling from the user; the data points are grouped, e.g., to minimize their mean distance within the cluster, while maximizing the distance between clusters. Another example of unsupervised learning problems is that of dimensionality reduction where the raw data consists of vectors in high-dimensional spaces, but only a subset of coordinates carries useful information with respect to a specific aspect. Being able to reduce the dimensionality of a vector of data is useful as geometric volume scales exponentially with the dimension of the space it is in, and so does the number of points needed to capture (learn) specific properties of an $n$-dimensional object.

Reinforcement learning. In reinforcement learning the learning agent (or algorithm) learns through the interaction with a target *task environment* and determines actions that maximize some established long-term objective. Rewards and/or penalties are dispensed for outputs that lead the agent closer to or farther from the objective, respectively. Reinforcement learning can have supervised or unsupervised settings, but it is fairly unrestrictive as, for instance, sub-optimal actions are usually not explicitly corrected.

Most of the selected experiments in this review belong to the category of supervised learning. Our scope in this work goes beyond reporting a series of studies where ML/QML algorithms have been successfully implemented to improve, e.g., the speed or accuracy of diamond-based quantum applications. Our intent is to provide a critical discussion of these ML/QML algorithms, detailing why and how they perform better then, e.g., traditional methods based on statistical least-square curve fitting. When relevant, we also highlight distinctive traits of ML/QML that show potential for being generalized to experiments beyond quantum metrology.

## 2.3 Quantum machine learning and machine learning

Quantum machine learning (or quantum learning) is an umbrella term that can designate different approaches, mainly: *i)* using machine learning to improve the measurement of a quantum process or *ii)* using intrinsically quantum systems applied to classical ML strategies to execute a task more efficiently.[22,31] In this context, improvement and efficiency are specific to the process or task of interest, but often consist in a speedup in execution or increased resolution/accuracy based on definite statistical estimators. We also note that, as we will see (e.g., cf. § 3.5), approach *(ii)* can help identify fundamental boundaries between classical and quantum schemes, with the prospect that quantum paradigms might simplify problems that are hard for classical ML approaches. In this review, we will show examples of both *(i)* and *(ii)*, but we will reserve the term QML to processes of type *(ii)* that exploit resources that are fundamentally quantum and are applied in a ML context. Processes of type *(i)* will instead be classified under the label ML as they pertain to demonstrations that make use of fundamentally classical ML approaches yet applied to a quantum-based measurement. A more exhaustive discussion that concerns QML and include further sub-classifications can be found in dedicated reviews.[31,37]

## 3. Machine and quantum learning in diamond-based quantum science

In the last few decades, diamond has emerged as a prime hardware material for quantum technologies (Fig. 1). Diamond is host to a large variety of color centers—isolated atom-like defects consisting of foreign atoms and vacancy complexes that display unique, addressable spin-optical quantum properties with long coherence time, room temperature operation and the ability to create entangled states.[23] The most notable of these are the nitrogen-vacancy (NV)[38] and the group IV color centers (Fig. 1a).[39] Owing to their properties, diamond color centers have been utilized in several realizations—both fundamental and practical—in various fields ranging from quantum communication and computation to quantum simulation, metrology and sensing (Fig. 1b–d).[40–51] The following is a series of demonstrations we selected to highlight the role machine and quantum learning are playing to further advance these fields and their relevant applications. Table 1 gives an overview of the main areas where ML and QML have been successfully applied to quantum-related technologies. For context, in the table, we separate applications that have been realized in diamond (blue) from those that have been realized in other systems/materials (dark orange).

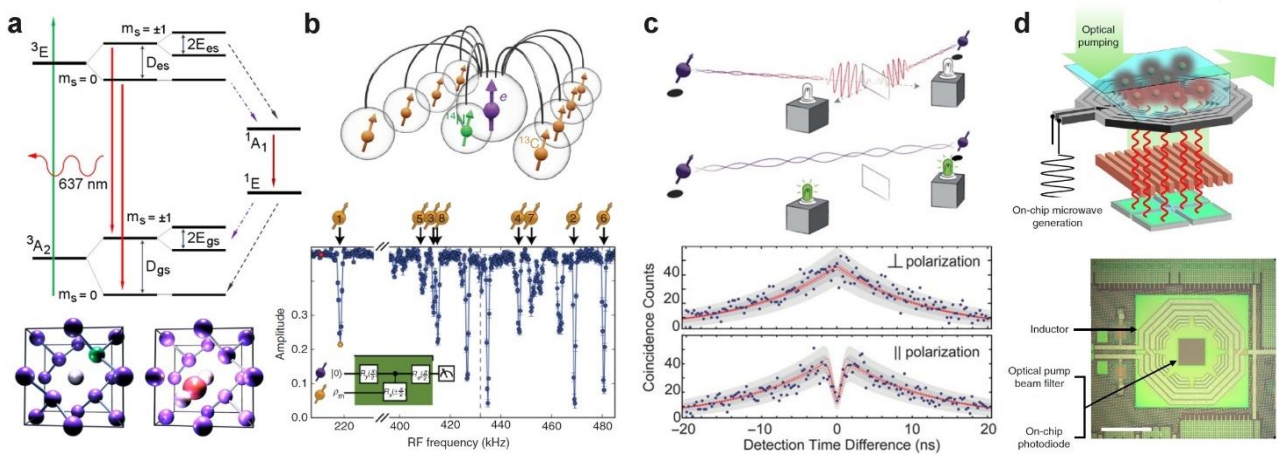

**Figure 1.** Diamond color centers and their quantum applications. **a)** Simplified energy level scheme of the nitrogen-vacancy (NV) center (top); crystalline structure of the NV center (bottom left) and of group IV color centers (bottom right) in diamond. **b)** Schematic of an NV electron spin acting as a central qubit connected by two-qubit gates to the intrinsic $^{14}$N nuclear spin and further eight $^{13}$C nuclear spins surrounding the NV center (top); corresponding nuclear spin spectroscopy (bottom) showing the frequency signals due to the interactions with the eight $^{13}$C nuclear spins. Reproduced with permission;[42] copyright 2019, American Physical Society.

**c)** Schematic for generating entanglement between distant spins (top): two photons, each entangled with a different spin, interfere on a beam splitter. Detection of a photon on each output projects the spins onto a long-distance entangled state. Two photon quantum interference measurement (bottom). Photons from different NV centers arrive onto a 50/50 beam splitter; coincidences are shown as a function of time between detection events on the two outputs. When the photons have perpendicular polarization no interference occurs, when they have parallel polarizations a pronounced suppression of coincidences near zero delay is visible indicating high contrast (~60%) two-photon interference and some degree of indistinguishability between them. Adapted and reproduced with permission.[46,52] Copyright 2013, Springer Nature and 2012, American Physical Society. **d)** CMOS-integrated quantum sensing chip (top). A green pump laser excites an NV ensemble in the diamond slab, while microwave fields manipulate NV electron spins through an on-chip inductor. A metal/dielectric grating absorbs the green pump beam and transmits the NV spin-dependent fluorescence to the on-chip photodiode. Top-view micrographs of the fabricated CMOS chip (bottom). Adapted and reprinted with permission;[51] Copyright 2019, Springer Nature.

Before delving into the specifics, we draw attention to a few general features ML and QML possess that can be harnessed to complement and/or improve quantum realizations—as we will see, mostly resulting in measurement speedups or increased accuracy. Above all, it is worth underscoring that both machine learning and traditional inference-based methods leverage statistical analysis of the data to make determinations. Yet, they do so somewhat differently. In traditional approaches based on least-square data fitting, a *complete* dataset is acquired for a specific measurement and a decision is made on the statistical analysis of such a set. The term 'complete' here indicates that the dataset is large enough to be fitted with a model, within a target level of accuracy (e.g. an acceptable error). Conversely, in ML the acquired dataset is often *sparse*; that is the information of one dataset might be incomplete—for instance because the integration time of the measurement is short. Yet, the sparse set still contains information: patterns that can be correlated to specific outcomes. Correlating patterns to outcomes is carried out by comparing the unknown sparse dataset to a large number of training sets (~hundreds or thousands) of known, previously-acquired data. Thus, machine learning has the large overhead cost of relying on the acquisition of several training sets to start, but the payoff comes in the ability to then being able to make predictions about an unknown dataset, even when this is sparse (i.e. there is not enough data for a full traditional least-square fitting analysis). This difference, emphasized by the expressions *long data* (relevant to inferential statistics) vs. *wide data* (relevant to ML), is key.[32]

Also, in inferential statistics, a model with unknown parameters for the mechanism governing the observed property is proposed and then correlations and statistical tools help determine the values of said parameters (note that if the assumptions about the model are wrong, the computed parameters might carry little to no meaning). Conversely, Machine Learning involves *learning methods* that aim at identifying correlations in datasets without prior determination of an underlying model for the mechanism under study. There is not an explicit formula for the distribution of the data, instead, ML algorithms make predictions in response to identifying patterns through general-purpose (learning) algorithms. In this review we will highlight how this ability of ML/QML to extract information from sparse datasets can result in significant improvements in measurement speed and accuracy for several quantum applications.

**Table 1.** Application of ML/QML to quantum technologies

|  | **Quantum characterization** | **Quantum measurements** | **Material/device design** |
|---|---|---|---|
| **Application** | o SPE characterization[53]<br>o Light sources classification[54]<br>o Light-based decision logic[55]<br>o Boson sampling validation[56,57] | o Magnetometry[58]<br>o Spin-state readout[59,60]<br>o Quantum clustering[61]<br>o Quantum tomography[62–64]<br>o Quantum phase estimation[65]<br>o Quantum experiment design[66,67]<br>o Quantum algorithm simulation[68] | o Material tailoring[9,69,70]<br>o SPE/spin defect stability[71]<br>o Devices design[72,73,19,74–77]<br>o Devices automation[18,78,79] |
| **Algorithms** | o Classical ML classifiers[80–84]<br>Regression, random forest, tree-based models, k-nearest neighbors, support vector machine, etc.<br>o Neural network classifiers[85] | o ML regression[33,37,86–90]<br>Linear, logistic, bayesian, support vector machine, etc.<br>o Hamiltonian learning[59,91]<br>o Quantum learning[21,31,37,92]<br>o Neural network regressors[85] | o Neural networks[93]<br>o Bayesian inference[89,90]<br>o Active learning[94]<br>o Reinforcement learning[95] |

Blue text indicates quantum-related applications realized in diamond while dark orange text those realized in systems/materials other than diamond

*3.1 Single-photon source characterization*

In many quantum photonic and sensing applications based on diamond color centers, one of the key requirements is the identification of bright, photostable single-photon emitters (SPEs) with high emission rates, high quantum yield, and desirable spectral properties such as narrow optical linewidth and high photon purity. The full characterization of SPEs can be a lengthy process as some measurements require relatively long integration times, making it a critical problem especially for applications involving a large number of quantum optical systems. For instance, the standard way of discriminating between emission from single emitters and ensembles requires carrying out a second-order autocorrelation measurement $g^{(2)}(\tau)$ between pairs of consecutive photons in a Hanbury-Brown-Twiss (HBT) interferometer. This is experimentally onerous since: *i)* the determination requires the statistical analysis (e.g. using the Levenberg–Marquardt or damped least-squares fit)[96,97] of a sufficiently populated histogram to reliably retrieve the $g^{(2)}(\tau = 0)$ value, and *ii)* the correlation rate is proportional to the square of the source intensity, which is relatively low for single-photon emitters.

This is a classification problem where the goal is to discriminate weather or not the autocorrelation function $g^{(2)}(\tau)$ *at* zero-delay time ($\tau = 0$) is lower than a threshold value $G$ ($G \leq 0.5$ conventionally indicates emission from single-photon emitters). Machine learning can solve such discrimination/classification-type problems rapidly and using much smaller data sets than those required in traditional approaches based on least-squares fitting of a model. Rapid discrimination of single-photon sources using ML and sparse datasets was recently demonstrated.[53] In the study, ML classifiers were used to conclude whether the photoluminescence from nanodiamonds containing nitrogen-vacancy (NV) centers originated from individual emitters or ensembles (Fig. 2a). The method yielded a classification accuracy of 95% within an integration time of less than a second—which is a ~100-fold speedup compared to the conventional Levenberg–Marquardt fitting approach. In addition to the speedup, another interesting aspect of the study was the way the ML classifier algorithms were trained and tested. Machine learning algorithms can predict outcomes only after being trained and tested with, in this case, sets of measured data from known, pre-characterized emitters. Here the authors also made use of emulated datasets. Nitrogen-vacancy centers can be modeled as three-level systems with both radiative and non-radiative transitions and associated rate equations (Fig. 1a). The authors could therefore model—numerically—the emission from NV centers and create an arbitrarily large number of emulated autocorrelation histograms from simulated HBT interferometry experiments, with any arbitrary level of uncorrelated background noise. This way of acquiring the training/test sets is noteworthy as it directly overcomes one of the biggest limitations of ML approaches—namely the need to gather, upfront, large sets of data to train and test the algorithm. It is important to note that this is possible only if the physics and dynamics of the system under study are well understood. Yet, it also underscores the vast potential of hybrid approaches where both ML and knowledge-driven modeling are combined to achieve high optimization efficiencies.[98]

In the study, the authors specifically used both experimental and emulated $g^{(2)}(\tau)$ histograms (~forty thousand) to effectively and rapidly quantify the accuracy of four different supervised ML classifiers (cf. §§ 4.1–4): *support vector classification* (SVC),[86] *gradient boosting classifier* (GBC),[82] *voting classifier* (VC),[83] and *convolutional neural network* (CNN).[85] Of these, CNN and VC were found to be the best performers (Fig. 2a). These ML-assisted classifiers could predict the 'single/not-single' nature of real emitters with 1-$s$ acquisition time sparse datasets and achieve a level of accuracy greater than 90%. Note that in a standard autocorrelation measurement a 1-$s$ acquisition histogram can have—depending on the brightness of the emitter(s)—as low as $N < 1$ average recorded events per bin, which is extremely low and not large enough to perform a traditional statistical analysis. For context, the prediction using the conventional Levenberg–Marquardt fitting method applied to these sparse datasets performed, expectedly, no better than a random guess. At least two orders of magnitude longer collection times (~mins, or $10 < N < 20$ average events per bin) were needed for the Levenberg–Marquardt fit to reach an accuracy comparable to that of the ML strategies. These numbers show the remarkable potential of ML approaches to significantly speed up specific measurements. This is especially true in this demonstration, for the training and testing of the supervised model was carried out, efficiently, through numerically simulated datasets, rather than solely on experimental ones. This eliminates most of the upfront costs typical of many supervised ML approaches that require large sets of known (usually experimental) data for the training and testing of the algorithm.

Shortly prior to this demonstration, machine learning had also been used in the related, more general task of differentiating between coherent and thermal photon sources.[54] This is another classification-type problem where machine learning can be very efficient. The probabilities of finding $n$ photons in coherent and thermal light are $P_{coh}(n) = e^{-\bar{n}}(\bar{n}/n!)$, and $P_{th}(n) = \bar{n}^n/(\bar{n}+1)^{n+1}$, respectively, where $\bar{n}$ is the mean number of photons in the beam. The photon statistics of thermal light displays random intensity fluctuations with a variance greater than the mean number of photons and a maximum always at vacuum. Conversely for coherent light, the maximum photon-number probability is $\sim n$. While fundamentally different, when the mean photon number is low the photon number distributions for thermal and coherent light become similar. Discrimination between the two sources is traditionally performed through statistical analysis of millions of measurements. In the study however, through the use of naive Bayes and ADAptive LINear Element (ADALINE) classifiers (cf. § 4.4) the authors demonstrated the ability to discriminate between thermal and coherent sources with high accuracy (61–90% for a corresponding number of data points 10–160) for an average number of photons $\bar{n} = 0.4$, i.e. well below 1.

The effectiveness of ML for such classification-type problems can find immediate practical application in a vast range of realizations ranging from imaging and remote sensing/metrology at extremely low light levels[99,100] to super-resolution microscopy based on single-photon autocorrelation measurements,[101,102] and assembly of scalable quantum photonic circuits requiring tens of single-photon emitters.[43] Additionally, we should remark that these ML-driven classification-type problems can be solved in a fully-automated fashion. This is key to achieve the technological and practical goal of engineering devices whose design, characterization and testing require automated, fast and large-scale production.[103]

*3.2 Single-photon decision making*

Tangential to machine learning, NV centers in diamond have also been used in a proof-of-concept demonstration about adaptive, autonomous decision making.[55] The study focused on harnessing the quantum nature of a single-photon NV emitter to engineer an artificial decision-making machine that could autonomously tackle problems such as the exploration-exploitation dilemma[104] and the multi-armed bandit problem.[105,106] These are traditional problems in probability theory and machine learning, where the goal is to make decisions between competing choices that maximize reward, while having limited knowledge and/or resources available. The simplest example of one such decision-making problem is that of a player who must select between two slot machines, $L$ or $R$, to maximize winnings. The reward probability of the slot machines are $P_L$ and $P_R$, respectively and may change over time. The player may test the machines to know which one is better at different times, while understanding that too much testing may result in excessive loss.

In the experiment, single photons from an NV center are sent through a polarizer, a half-wave plate and a polarizing beam splitter that separates the incoming photons—based on their horizontal/vertical polarization—to two separate channels. The half-wave plate controls the discrimination process: when at 45° the photons have a 50% chance to either go to the left, $L$-channel, or to the right, $R$-channel. As the photon is detected in a channel, a corresponding reward is dispensed with probability $P_L$ or $P_R$. If the reward is dispensed (retained), a motor automatically rotates the half-wave plate slightly off the 45° position to favor the next photon to land on the same (opposite) channel. By iterating the process, the rotating half-wave plate autonomously makes the correct decision: i.e., it chooses the slot machine with a higher reward probability. In the study, the authors also test more elaborate decision-making schemes and the adaptability of the machine by changing the reward probability after several cycles. As expected, the machine displays a tradeoff between reaching the correct answer quickly and maximizing the correct selection rate.[107]

In most demonstrations, machine learning approaches are discussed in the context of achieving better performances—e.g. higher speeds and accuracies—than traditional methods based on statistical inference. This work, on the other hand, has the merit to showcase how the quantum nature of single-photon generation and detection can be harnessed to engineer autonomous, intelligent machines purely based on a physical mechanism. The realization of a decision-making scheme based primarily on light and its attributes also helps broaden the scope of quantum information science beyond its current focus of developing hardware and protocols for quantum computing and quantum key distribution.

*3.3 Magnetometry*

The nitrogen-vacancy (NV) center in diamond has become an archetypical system for applications in quantum sensing and metrology, with both fundamental and practical realizations ranging from nanoscale magnetometry[108–110] and electrometry[111] to decoherence microscopy,[112] superresolution imaging,[113] thermometry,[114–116] optical trapping[117,118] and Forster resonance energy transfer.[119] The NV center possesses unique spin-optical properties that match those of isolated atomic systems, whilst also displaying relatively long coherence times at room-temperature and the technological and logistic advantages of a solid-state host.[38] The center is a spin-1 system with a triplet ground and excited states (Fig. 1a). A combination of optical and resonant microwave excitation allows for the initialization, coherent manipulation and readout of the center's spin at room temperature, through a spin-dependent intersystem crossing mechanism that produces both optical spin initialization of the NV's ground state and spin-state-dependent fluorescence contrast for its optical readout. Additionally, the center's spin states are sensitive to magnetic and electric fields, strain, and temperature variations, allowing to map any of these external stimuli directly onto said states. Amongst the various sensing schemes, NV-based magnetometry is arguably one of the best established, for which numerous strategies have been developed to maximize both spatial and signal resolution.[26,120] The bases to NV dc-magnetometry are Zeeman splitting of the $m_s = \pm 1$ spin states and/or Ramsey interferometry. Briefly, the NV is initialized in the $m_s = 0$ ground state via laser excitation, while application of a microwave pulse resonant with the $m_s = 0 \leftrightarrow \pm 1$ transition creates a superposition between the two corresponding spin states. The intensity and direction of the magnetic field are then measured as a shift in the resonant microwave frequency or as an accumulation of the relative phase between eigenstates, depending on the measuring scheme. For the scheme based on Ramsey interferometry, the phase accumulation $\phi = 2\pi f_B \tau$ for a Larmour frequency $f_B = \gamma B / 2\pi$ is directly proportional to the magnetic field $B$ to be measured ($\gamma$ is the electron gyromagnetic ratio of the magnetic moment to the angular momentum) and to the time $\tau$ between the two consecutive $\pi/2$ microwave superposition and readout pulses. Collecting statistics for a series of different values of $\tau$ produces a fringe of phase varying with time, from which the magnetic field $B$ can then be extracted.[121–123] The resolution on the measurement of $B$ is limited by quantum projection noise and, ultimately, by the spin coherence time $T_2^*$,[108] which sets the Heisenberg limit (HL).[124] The quantum projection noise can be reduced and the Heisenberg limit reached through repeated measurements, at the cost of longer integration times. Notably, at cryogenic temperature the spin states can be accessed such that fluorescence is absent from the $m_s = \pm 1$ state—resulting in a stronger optical differentiation (and hence higher signal resolution) between the $m_s = 0$ and $m_s = \pm 1$ states. However, at room temperature, the unresolved spin optical transitions reduce the fluorescence contrast, requiring several repetitions of the Ramsey cycle per value of $\tau$, reducing the resolution of the technique by several orders of magnitude.

Numerous strategies have been developed to overcome the resolution limit at room temperature and, amongst these, methods based on ML are particularly appealing. The first demonstration of this was realized in 2019.[58] In the work, the authors used Bayesian phase estimation and Hamiltonian learning techniques (cf. §§ 4.6, 4.7)[59,59,125] to develop a so-called magnetic-field learning (MFL) algorithm that could achieve an average one-photon readout from a single NV center at room temperature and produce a level of sensitivity comparable to that of cryogenic operation. The MFL algorithm works in steps called epochs (Fig. 2b). During each epoch a Ramsey measurement is performed with a certain accumulation time $\tau_i$ and the measurement outcome is used to set the subsequent accumulation time $\tau_{i+1} \cong 1/\sigma_i$—determined as the prior distribution is updated $P_i \to P_{i+1}$ through Bayesian inference ($\sigma_i^2$ is the uncertainty embedded in the prior $P_i$). This differs from traditional approaches where several Ramsey cycles are acquired per each accumulation time $\tau_i$, as $\tau_i$ is varied from a minimum to a maximum value in an unguided fashion. The performance of the MFL algorithm can be measured and compared to traditional approaches by monitoring the precision $\eta^2 = \sigma_B^2 T$, defined as the product between $\sigma_B$ (the uncertainty on the measurement of the magnetic field $B$) and $T$ (the total acquisition time $T = \sum_i^N \tau_i$ where $N$ is the total number of epochs). Note that the number of MFL epochs coincides with the number of Ramsey cycles, thus allowing for a direct comparison with the acquisition time of standard methods. For completeness, in the study a second definition of $T$ as the total acquisition time including overhead and computation time was also considered and again compared to the equivalent of traditional approaches. In the work, the authors show that the precision $\eta^2$ scales as $T^{-1}$ for the MFL algorithm vs. the $T^0$ scaling of other standard methods based on fast Fourier transform (FFT). More specifically, the authors show that the MFL algorithm performs better than alternative methods regardless of the chosen metric—uncertainty, $\sigma_B$, or total acquisition time, $T$. Importantly, they also show that, at room temperature and with as little as 1 detected photon, they reach a resolution of $60\ \mathrm{nTHz^{-1/2}}$ in $\sim 10\ \mathrm{ms}$ from a total of $4000$ individual Ramsey sequences. This directly compares with performing the measurement under cryogenic conditions where single-photon readout is possible.[126] Interestingly, the MFL algorithm is multiparametric—which is typical for machine learning approaches—i.e. it can make predictions through minimizing the joint uncertainty on several features, rather than a single parameter. This allowed the authors of the study to switch between two modes: a first one where MFL uses $B$ as the only learning feature and a second one, at a later stage, where the algorithm uses both $B$ and $T_2^*$. The benefit is that by estimating $B$ and $T_2^*$ simultaneously the algorithm can avoid suggesting values of $\tau_i$ for the following iteration, greater than $T_2^*$ as these carry no useful information. Note that instead, in traditional approaches, $T_2^*$ is estimated independently prior to the sensing experiment, dramatically increasing the total running time of the measurement. To showcase the remarkable performance of the MFL algorithm, the authors demonstrate rapid tracking of instantaneous changes in the field $B$, showing that the algorithm quickly ($\sim 10$ epochs) converges to the new set value for $B$, much faster ($\gtrsim 2\times$, from the data shown) than traditional approaches (Fig. 2b).

As per the case of ML being used to rapidly discriminate between 'single' and 'not-single' emitter (cf. § 3.1), the speedup of the measurement is not simply a matter of practical convenience. As shown in this study, the speedup translates directly into an increase in resolution of the magnetic field sensing technique and in the ability to reproduce at room temperature performances otherwise only achievable at cryogenic temperatures. This matters as it makes the sensing method available to a series of investigations—such as those of biological environments—incompatible with low-temperature experiments.

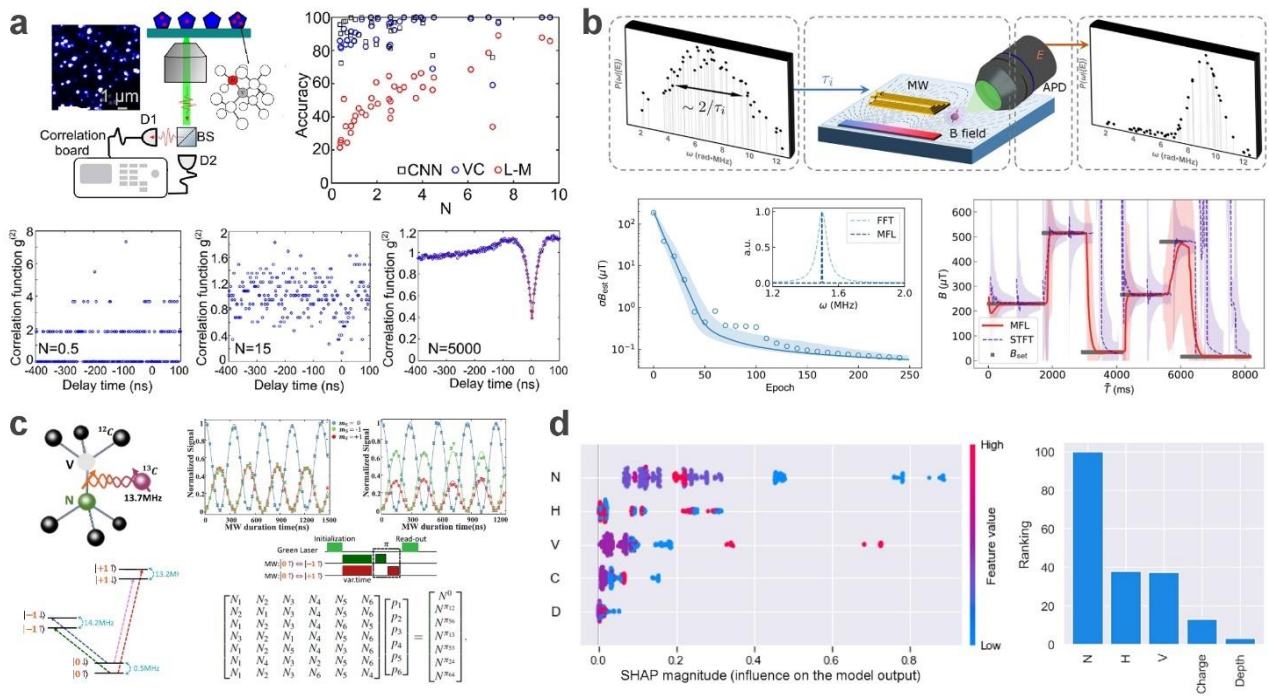

**Figure 2.** Diamond-based quantum applications exploiting machine and quantum machine learning. **a)** Hanbury-Brown and Twiss interferometer (top left) to distinguish single NV centers from ensembles in nanodiamonds. Corresponding second order autocorrelation measurements $g^{(2)}(\tau = 0)$ at three different

integration times (bottom) such that the average number of counts $N$ per bin is 0.5, 15 and 5000, respectively. Graph showing the accuracy of predicting NV centers being 'single'/'not single' as a function of the average number of counts $N$ per bin (top right), using two machine learning methods, CNN (black squares) and VC (blue circles), and using a traditional Levenberg-Marquardt curve fitting approach (red diamonds). Figure adapted and reproduced with permission;[53] copyright 2020, John Wiley & Sons. **b)** Epoch of the magnetic-field learning algorithm (top) where the uncertainty encoded in the prior distribution $P_{i-1}$ determines the phase accumulation time $\tau_i$ for the next set of Ramsey sequences; the outcomes $E$ from these sequences are measured and the prior distribution updated $P_i \rightarrow P_{i+1}$ through Bayesian inference. Experimental results for scaling of precision (bottom left): the estimated uncertainty $\sigma(B_{est})$ is plotted as a function of the epoch number; data from one sample run is shown as blue circles. The inset shows a plot of the final $\sigma(\omega_{est})$ in the Ramsey frequency for a typical protocol run, using FFT and MFL. Experimental magnetic field tracking (bottom right); step changes in $B$ are indicated by the gray bars, the solid red line represents the typical performance of MFL, while the dashed purple line indicates the outcome of a traditional short-time Fourier transform protocol. Figures reproduced with permission;[58] copyright 2019, American Physical Society. **c)** Structure and energy level scheme of the electron nuclear spin system used in the work (top and bottom left). Experimental generation of an arbitrary three-level superposition state (top right): population vs MW duration for $\Omega_{m_s=+1}/\Omega_{m_s=-1} = 1$ (left) and $\Omega_{m_s=+1}/\Omega_{m_s=-1} = 1/\sqrt{2}$ where the ratio of population $P_{m_s=+1}/P_{m_s=-1}$ is equal to the square of that of the MW amplitude $\left|\Omega_{m_s=+1}/\Omega_{m_s=-1}\right|^2$. State population determination (bottom right): after getting all the single-level PL rates $N_i$, the population of the final state is measured, $\text{diag}(\rho_f) = (p_1, p_2, ..., p_6)$, by flipping populations between different levels in the final state; $\pi_{ij}$ denotes a selective $\pi$ pulse between levels $i$ and $j$, and $N_0$ represents the PL rate of the final state without any flipping. Figures adapted and reproduced with permission;[61] copyright 2020, American Physical Society. **d)** Beeswarm plot (left) of the absolute Shapley values for each feature of each instance in the testing set and the feature important profile (right) showing the relative importance of different features—number of nitrogen atoms per defect, number of vacancies, the presence of hydrogen at the surface, the defect's charge and its distance from the surface—in determining the likelihood of a defect configuration to occur. Figures Adapted and reproduced with permission;[71] copyright 2022, Elsevier B.V.

*3.4 Spin dynamics monitoring*

Prior to being used in practical magnetometry applications (cf. § 3.3), Bayesian inference and Hamiltonian learning had been utilized for the more fundamental realization of monitoring the electron spin dynamics of a single NV center.[59] This demonstration directly showcases how ML can provide a practical solution to the fundamental problem that predicting real data from a model—in the exponentially-large configuration space of quantum systems—is generally intractable with classical computers. This is for instance the case when trying to capture the dynamical Hamiltonian evolution of a system, for learning the Hamiltonian relies on an estimation of likelihoods that are exponentially hard to compute classically. Quantum Hamiltonian learning (QHL) solves this issue through the use of machine learning and a quantum simulator, which makes the classically-intractable problem of learning the Hamiltonian tractable because the quantum simulation is exponentially faster than classical techniques.[127]

Briefly, given a real quantum system, the quantum simulator mimics its dynamics through a model Hamiltonian $\hat{H}$ parameterized with respect to some parameter(s) $x$, i.e. $\hat{H} = \hat{H}(x)$. A real measurement $D$ is performed and is used to estimate its associated likelihood $P(D|x)$, i.e. the probability that the experimentally measured value $D$ was produced by the parameters $x$. Each likelihood is then used to infer the posterior distribution of the parameters $x$ via Bayes' rule and to calculate the subsequent time-evolution step (cf. §§ 4.6, 4.7).

In this particular demonstration, the real system under study was the electron spin state of an NV center while the quantum simulator consisted of a silicon photonic device. The NV undergoes a single Rabi cycle, where the center is initialized in the $m_s = 0$ spin state by optical excitation, and then subjected to a pair of microwave pulses resonant with the $m_s = 0 \leftrightarrow -1$ state transition. The pulses are separated by a delay time $t$ and have fixed power; the time $t$ determines the evolution of the electron spin and the relative probability of measuring the system in either the $m_s = 0$ or $m_s = -1$ state (the two spin states are optically distinguishable due to the spin-selective fluorescence contrast, Fig. 1a). Rabi oscillations of the NV's electron spin can be modeled with the parameterized Hamiltonian $\hat{H}(f) = \hat{\sigma}_x f/2$, with $\sigma_x$ as the quantization axis and $f$ the Rabi frequency. The silicon-photonics chip simulates the parameterized model Hamiltonian $\hat{H}(f)$ to predict the Rabi frequency $f$ and allows for the calculation of the likelihood function $P(D|x)$, where in this case $x = f$ and $D$ is the measurement of the NV spin state after it was prepared in the state $|\psi\rangle$, let to evolve for a time $t$ and measured in a basis $\{|D\rangle\}$. At each iteration of the algorithm, the evolution time $t$ is chosen—adaptively—to maximize the spin undergoing a single Rabi nutation.

Similar to the discussed performances of single-emitter classification (cf. § 3.1) and magnetic sensing (cf. § 3.3), the ML-based approach converges to the real Rabi frequency of the system much faster (~50 vs. ~200 measurements) than performing the statistical analysis on a complete dataset where the delay time $t$ is varied in an unguided manner from a minimum to a maximum value, and integration is performed over several repetitions per each value of $t$. Another advantage of the Bayesian framework used in the study is that one can directly compute the variance from the posterior distribution and identify potential limitations in the model. The authors also show that by introducing a parametrized Hamiltonian of higher complexity, a time-dependent Hamiltonian with chirping, $\hat{H} = \hat{H}'(f, \alpha; t) = \hat{\sigma}_x(f + \alpha t)/2$, they could further decrease the variance (therefore increasing resolution) of the predicted Rabi frequency.

Once again, the demonstration underscores how machine learning approaches are not merely advantageous for a speedup of the measurement. They can also guide the development of the model itself—through, e.g., a practically beneficial improvement in measurement resolution—and potentially reveal nuances of the physical processes under study.

More recently, another machine learning approach based on multi-feature linear regression was used to improve the readout accuracy of the electron spin state of an NV center.[60] As already discussed (cf. § 3.3), the electron spin state of a NV center can be initialized ($m_s = 0$) by optical excitation, manipulated ($m_s = 0 \leftrightarrow \pm 1$) through application of a resonant microwave field, and read out optically due to the difference in scattered photons (~30%) between the excitation-relaxation cycles of the $m_s = 0$ and $m_s = \pm 1$ states (Fig. 1a). Experimentally, the different spin states can be distinguished by measuring on a photodetector the difference in fluorescence during a time-gated window of specified duration $\Delta t$. Note that in general the fluorescence is an integral over $\sim 10^5 - 10^9$ repetitions of the measurement, which allows for detecting sufficient signal contrast. Said contrast depends on several factors including power fluctuation of the excitation laser, sample drift, environmental effects, etc. In this work, the authors focus on developing a ML-based method to specifically reduce shot noise, which—while seemingly small in scope—has the appeal of advancing a low-cost solution, resource-wise, that is widely applicable.

In traditional approaches, the aim is to find the optimal duration $\Delta t$ of the gated time window that produces the highest contrast $C$ and the lowest variance $V$ between the fluorescence signals of the $m_s = 0$ and $m_s = \pm 1$ states (note that the $\Delta t = \Delta t_C$ that maximizes $C$ might not coincide with the $\Delta t = \Delta t_V$ that minimizes $V$). The quantities $C$ and $V$ are defined as: $C = (L_0 - L_1)/L_0$ and $V = (L_0 + L_1)/2(L_0 - L_1)^2$, where $L_0 = \sum_i^N x_i^{(0)}$ and $L_1 = \sum_i^N x_i^{(1)}$, and $x_i^{(0)}$ and $x_i^{(1)}$ are the number of photon counts recorded in each $i$-th bin ($i = 1 - N$) the time interval $\Delta t$ is divided into (the superscripts

'0' and '1' refer to the two cases of measuring the fluorescence from the prepared $m_s = 0$ and $m_s = \pm 1$ states, respectively). The ML approach proposed in this study adopts a different strategy. Rather than considering the fluorescence signals as trivial sums, $L_0$ and $L_1$, of the counts $x_i^{(0)}$ and $x_i^{(1)}$ in each $i$-th bin, the ML algorithm treats the fluorescence $p$ as a function that is a weighted sum of the counts in the bins: $p^{(0/\pm1)} = \sum_i^N a_i x_i^{(0/\pm1)} + b$ ($b$ is a constant). Briefly, counts $x_i^{(0)}$ and $x_i^{(1)}$ in each $i$-th bin are not weighted equally; they are instead each multiplied by an individual weight $a_i$ that is optimized (alongside the number of time bins $N$) to yield the best $C$ and $V$. Collaterally, the ML algorithm is not restricted to just using data within the time window defined by the gate and utilizes instead all the data in the sets: the ML algorithm simply weighs certain bins more or less, regardless of the cut-off time. The weights are determined from known training data, as being those which minimize a least squares error (LSE) cost function that includes two terms: one for the contrast $C$ (desirably high) and one for the variance $V$ (desirably low). By using this simple approach on a set of Rabi data, the authors were able to show 7–44% higher accuracy in assigning the correct spin state of the NV electron, where the difference in accuracy improvement depends on whether the ML algorithm was compared to traditional methods using a $\Delta t$ optimized to minimize $V$ or to maximize $C$ (again, in general, $\Delta t_C \neq \Delta t_V$).

The ML multi-regression model proposed in this study showcases a few key features. It is a relatively simple method and even if the improvements in accuracy were more modest, it would be of interest given the low requirements in terms of experimental/computational resources. It is widely applicable to any fluorescence-based measurement involving color centers in diamond or—universally—to any experiment where the measured signal is digital, and the discrete bins have relative intensities that correlate to specific values of an observable of interest. Finally, the proposed approach highlights that there are specific sets of problems where ML offers remarkable advantages over traditional methods based on statistical least-square fitting.

While not explicitly touched upon in this work, ML methods can determine parameters of interest with desired accuracy using a—usually significantly small—fraction of the data required by traditional methods based on least square curve fitting. This is because the latter might need many data points for the fitting curve to 'reach convergence,' while the former is based on pattern-recognition strategies that can determine the value of a variable from just a few strong correlations built upon known training sets of data. The tradeoff is that ML requires at least one—usually several—sets of known training datasets from which said patterns and correlations can be established before the model can be used to make determinations about unknown data, with a target level of accuracy. This highlights, again, the conceptual difference between traditional statistical inference, which relies on *long data*, and ML strategies, which instead rely on *wide data*.

*3.5 Quantum classification and clustering*
A vast proportion of machine learning problems require manipulating and classifying data sets that are large both in size and dimension—data is often represented as vectors in high-dimensional spaces. As already alluded to earlier (cf. § 3.4), classical algorithms are inefficient for solving such problems, for their computation time scales polynomially with the number of vectors and the dimension of the space (see below). Conversely, quantum machines are good at manipulating high-dimensional vectors in large tensor product spaces and can offer an asymptotically exponential speedup in computation, especially when dealing with sparse, high-rank matrices. In light of this, quantum learning approaches have been advanced for solving classification and clustering problems where the aim is to rapidly and accurately assign objects (effectively points or vectors in multi-dimensional space) to existing groups (clusters).[128–130] Note that according to the definitions we used (cf. § 2.3), such approaches fall under the banner of quantum machine learning, for intrinsically-quantum systems or protocols are incorporated into classical ML strategies to execute a task more efficiently or more rapidly.

In this context, recently the NV center in diamond has been utilized to test one such quantum machine learning clustering approach.[61] In the study, the authors specifically use the electron spin ($S = 1$) of an NV center coupled to the nuclear spin ($S = 1/2$) of a proximal $^{13}$C. The $^{13}$C nuclear spin ($|\uparrow\rangle$ or $|\downarrow\rangle$) is used to encode the vectors, while the NV electron spin-qutrit ($|0\rangle$, $|+1\rangle$, $|-1\rangle$) is used as the ancilla to perform the projection measurement and readout (Fig. 2c). The hyperfine coupling between the $^{13}$C and NV allows using a relevant combination of radiofrequency (rf) and microwave (MW) pulses to prepare any arbitrary electron-nuclear entangled state. In the study, the authors' goal is to prepare a set of such arbitrarily created test vectors $\vec{u}$ and assign them to one of two possible states. The assignment is carried out through a maximum likelihood approach based upon estimating the inner product and shortest distance between $\vec{u}$ and each of the two target quantum states (clusters).

To fully characterize the prepared test entangled state, quantum state tomography is needed, which for the NV-$^{13}$C system would require measuring 36 basis states to reconstruct the density matrix of this $3 \times 2$-dimensional system. However, the quantum classification algorithm can achieve that by only estimating the diagonal element of the density matrix, making the process significantly more efficient. Note that the NV-$^{13}$C system is ideal for this proof-of-principle demonstration as, experimentally: *i)* any test entangled state can be produced from the initialized $|0\uparrow\rangle$ state upon controlling the amplitude, frequency and duration of the rf and MW pulses, and *ii)* the projection measurement and readout can be done optically by measuring the relative fluorescence of the test state with respect to that of the $m_s = 0$ electron (the ancilla) spin state. For completeness, the readout requires prior calibration of the photoluminescence rates of the different levels by pumping the initial $|0\uparrow\rangle$ state to each level, and measuring the corresponding fluorescence intensity, but this can be readily carried out. Also, in the experiment the authors had to design a specific scheme for optical, rf and MW excitation to guarantee that the entire state preparation and projection process could take place within the coherence time $T_2^*$ of the NV electron spin.

Aside from the specific details though, proof-of-principle realizations such as this have a much more general appeal. They demonstrate that the problem of assigning $N$-dimensional vectors to one of several clusters of $M$ states can be computed by a quantum learning approach in exponentially less time than the fastest classical algorithm: $O(\log(MN))$ vs. $O(\text{poly}(MN))$, respectively.[128,129] This is relevant for many proposed future applications relying on analysis and encoding of 'big data' (either classical or quantum). To put this into perspective, currently the rate of electronic data generated per year is $\sim 10^{18}$ bits. This entire dataset could be represented by a quantum state using 60 bits ($10^{18} \approx 2^{60}$), and the clustering assignment could be performed using a few hundred operations. The entire information content of the universe is estimated to be $\sim 10^{30}$ bits, for which data representation and analysis could be carried out with 300 bits ($10^{30} \approx 2^{300}$)—well within the capacity of quantum learning algorithms run on a modest quantum computer.[128] At the same time, the benefits of quantum machine learning go beyond a mere speedup in computation. For instance—relevant to quantum information technologies—quantum machine learning provides an intrinsically higher level of privacy. Only $O(\log(MN))$ calls to the quantum database are required to perform the cluster assignment, while $O(MN)$ calls (the size of the database) would be needed to uncover the actual data. As a result, while the users can still perform the assignment, they only have access to an exponentially small fraction of the full database itself.

*3.6 Color centers synthesis and stability*

As briefly anticipated in the introduction, ML can be a powerful tool for material science and design owing to its ability to efficiently find patterns and trends, handle multi-dimensional and heterogeneous data and improve continuously with access to more observations—all while being intrinsically and fully automatable. Models and simulations based on ML approaches can, for instance, quickly narrow down the parameter space of specific variables involved in fabrication processes.[18,131,132] They can also be used to control, tweak or even design ad-hoc properties of materials,[69] heterostructures[9,70] and devices,[19,74,75] again, while being suitable for fabrication strategies that require large-scale, fast and automated production.

The potential of ML for diamond material science was recently showcased in a study assessing the stability of extrinsic defects in diamond.[71] As we have discussed, diamond is host to a wide range of defect complexes consisting of foreign atoms and/or vacancies in the crystalline matrix, which are the basis of several of the proposed diamond quantum science and technology applications. In the study, the author utilizes a ML learning approach based on neural networks to predict the stability of particular defect complexes in diamond—mainly nitrogen-vacancy (NV) related defects—as a function of parameters such as nearby concentration of other chemical species, charge state and surface depth and chemistry (Fig. 2d).

The ML algorithm used is a multi-layer perceptron (MLP), which is one of the most fundamental types of neural network architectures (cf. § 4.4). Briefly, the neural network determines the most likely output from a given set of inputs. Here the output is the likelihood of occurrence of a particular defect configuration (measured as relative defect energy), while the inputs are parameters such as the number of nitrogen atoms per defect, the number of vacancies, the presence of hydrogen at the surface, the defect's charge and its distance from the surface. The inputs and output are connected through a network of nodes (neurons) organized in layers. Each node is connected to every node in the previous and following layer with an associated weight and activation threshold that are determined by training the network with a known set of inputs and outputs (cf. § 4.4). Once trained, the network returns the most likely output for a given set of inputs, on the basis of the determined activation thresholds and weights. From a practical standpoint, the MLP algorithm leads to solving a system of equations that determines whether, for instance, a nitrogen-vacancy complex is more likely to form a negatively-charged $NV^-$ center rather than a neutral $NV^0$ or an isolated N atom, based on its (input) parameters. Notably, the algorithm can also predict the likelihood of formation of other defects not included in the training set, such as a divacancy complex, provided that the configuration can be described by an appropriate set of inputs.

In parallel, the author was also able to determine the contribution of every input characteristic to the occurrence of a target defect complex through the analysis of the Shapley values (Fig. 2d).[133] These represent the marginal contribution of an individual input feature of an individual instance to the predicted output or, in other words, how much each input feature affects the determination of a certain output. For instance, the Shapley analysis conducted in this study reveals that the amount of nitrogen is the most important factor in predicting the probability of forming an NV-related complex; the passivation of the surface is not as crucial as one might expect, and the depth below the surface has only a small influence on a defect being charged or neutral. This type of fully-computational analysis is useful, for the MLP neural network and Shapley values analysis give a quantitative measure of the input-output relation and trends. The analysis can thus help guide the choice of parameters for material synthesis, e.g. by narrowing the range of possible experimental values without actually having to test them during the material fabrication process itself. Besides its practical benefits, this type of analysis is also interesting from a fundamental point of view, for it can help complement the existing knowledge on the formation mechanisms of these nitrogen-related complexes in relation to, e.g., vacancy migration, particle size or overall nitrogen concentration.

Overall, the main merit of this study is to highlight how machine learning can be a powerful tool to complement and support discovery, design as well as process optimization in material science and device fabrication. Moreover, the approach is efficient and inexpensive as it does not actually require exploring the parameter space, fully, for all the variables involved in the fabrication process.

## 4. Machine Learning algorithms

In this section, we summarize a selection of the various ML/QML algorithms that have been used in the quantum-related applications we discussed in this review; they are introduced in the same order they appear in § 3. We present these methods from a general standpoint, without having any one specific application in mind. The section is meant as a reference that, in combination with § 3, aims at making the review a crossing point between the fields of quantum science and machine learning.

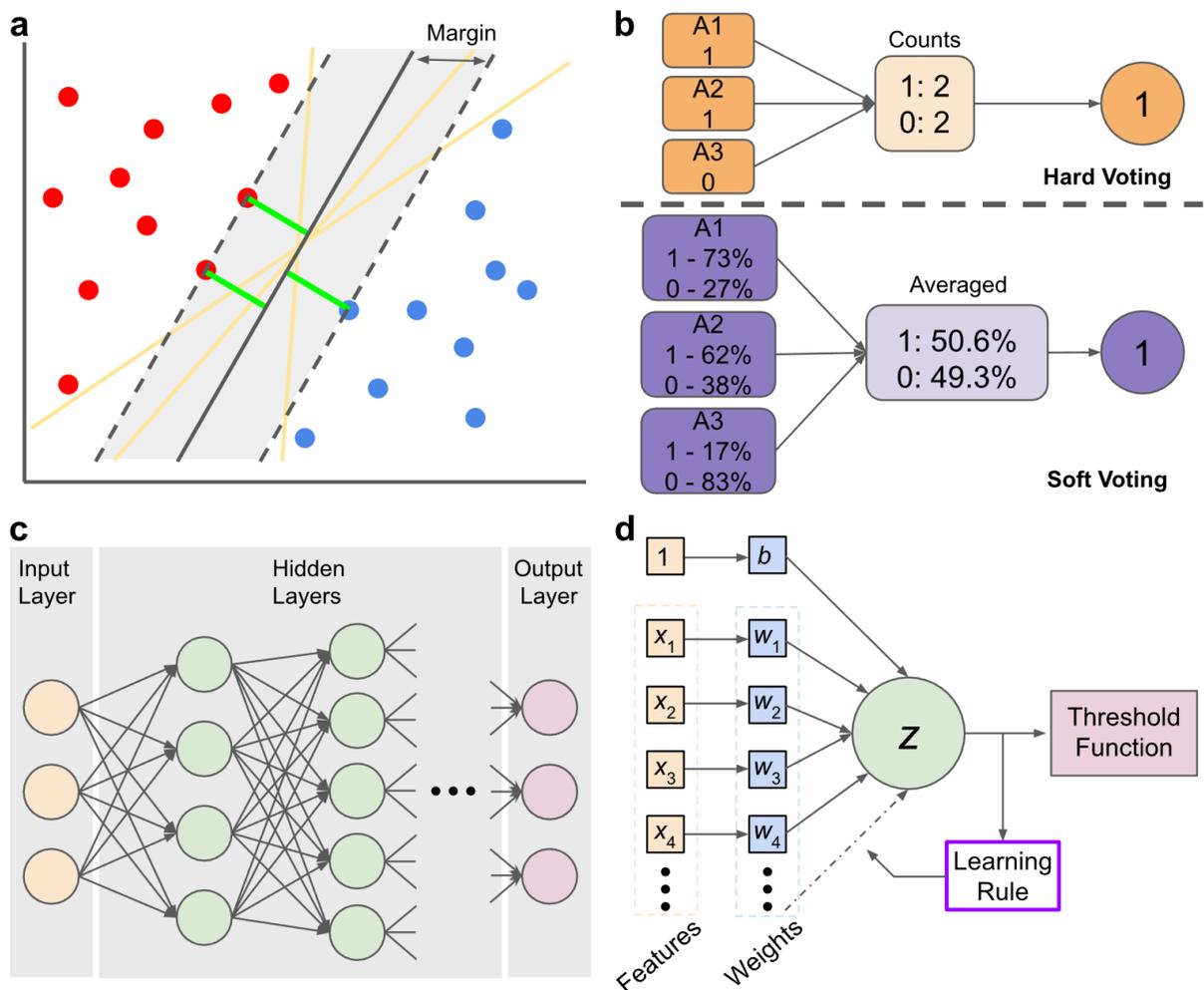

**Figure 3.** Machine learning algorithms. **a)** Support vector classification (SVC), structure. The red and blue points represent two different classes in a data set. The solid black and yellow lines show possible separating hyperplanes—in this case a line in two dimensions—with the black line representing the optimal separating hyperplane. The green lines show the distance between the support vectors and the hyperplane. The dashed black lines represent the boundary of the margin which is determined by the support vectors. **b)** Voting classifier (VC), structure. In hard voting, algorithms one through 3 output a single answer: the output that occurs most often is then chosen by the VC as the final answer. In soft voting, a weighted average of the probabilities is taken to determine the final output. Note: to simplify the figure all algorithms were given equal weighting. **c)** Deep Neural Network, structure. The network displays an input layer where the data enters the network, hidden layers that process/sort through the data, and an output layer that presents the results. **d)** ADALINE neural network, structure.

*4.1 Support vector classification*

Support Vector Classification (SVC) (also referred to as Support Vector Machines, SVM) is a *supervised*, *binary* classification technique that separates a set of data into two groups by finding the *optimal hyperplane* that divides them (Fig. 3a).[86,88] Hyperplanes are planes in a $(n-1)$ dimensional space—$n$ is the number of features—that set a decision boundary and separate the data points into two distinct classes. Note that SVC is natively binary; it can however treat multi-classification problems with more than two groups, by breaking them down into multiple binary classification problems.

Of all the possible hyperplanes, the optimal hyperplane is the one that displays the greatest *functional margin*—i.e. the largest distance between the hyperplane itself and the nearest data point(s) in each of the two groups on opposite sides of it. These nearest points define a set of $n$-dimensional *support vectors* that uniquely identify the *maximum-margin hyperplane*. A support vector classification algorithm is therefore a supervised, linear, maximum-margin classifier that aims at minimizing the generalization error—which determines how accurately the algorithm can predict outcomes of unknown data—by finding the largest distance between nearest training-data points belonging to opposite groups.

When implementing SVC, the training data is presented to the algorithm in the form of *instances*, i.e. features-label pairs $(x_i, y_i)$, $i = 1, 2, 3, \ldots, N$, where $x_i \in \mathbb{R}^n$, and $y \in \{-1, +1\}$.[84] Note that an instance in this context refers to a single row in the table (consisting of $N$ rows and $(n+1)$ columns) of $n$ features and 1 label. So for example, the $i$-th features-label pair refers to the vector $x_i$ with its $n$ features along with its corresponding label $y_i$ in the $i$-th row of the table. The resulting hyperplane takes on the form

$$h(x) = w \cdot x + b \qquad (1)$$

Where $w \in \mathbb{R}^n$ is the vector normal to the hyperplane and $b \in \mathbb{R}$ is a scalar value. The function $h(x)$ is such that $h(x) = 0$ for any point on the hyperplane. For a linearly separable data set, if the two classes are perfectly separable, all labels $y_i = -1$ have $h(x) < 0$, and all labels $y_i = +1$ have $h(x) > 0$.

In eq. 1, $w$ and $b$ are determined from the provided sets of training data by solving

$$\min_{w,b,\xi} \left\{ \frac{\|w\|^2}{2} + C \sum (\xi_i)^k \right\} \qquad (2)$$

Subject to the constraint

$$y_i(w^T \cdot x_i + b) \geq 1 - \xi_i, \quad \xi_i \geq 0 \, \forall \, x_i \qquad (3)$$

For all $x_i$ in the training set, where $C$ and $k$ are parameters associated with the cost of misclassification and $\xi_i$ is the *slack* variable (one for each $i$-th data point). The slack variable is introduced to account for the fact that not all data sets will be perfectly, linearly separable; $\xi_i$ thus determines the degree to which the constraint on the $i$-th datapoint can be violated (i.e. how much the $i$-th training point is allowed to be within the otherwise prohibited margin). Specifically, if $0 < \xi_i < 1$ then the point is within the margin, but on the correct side of the hyperplane, whereas if $\xi_i \geq 1$ the point is on the incorrect side of the hyperplane. The case of $\xi_i = 0$ indicates that the point was correctly classified, i.e. it is on the correct side of the hyperplane, on or beyond the margin. Note that through eq. 2, the aim is to simultaneously minimize the number of slack variables and their violation of the margin, while maximizing the margin itself (i.e. the separation between the two groups).

In the case where the data cannot be separated by a linear hyperplane with acceptable amounts of slack, the input vectors $x_i$ can be mapped to a higher dimensional space $\varphi(x_i)$ where a linear separating hyperplane can be found. This approach is known as the *kernel trick* and uses the *kernel function* defined by $K(x_i, x_j) \equiv \varphi(x_i)\varphi(x_j)$.[84,134] The same approach discussed above is then applied to the new instance-label pairs $(\varphi(x_i), y_i)$. The resulting hyperplane can then be mapped back to the original space resulting in a *non-linear separating hyperplane*.[84]

*4.2 Gradient boosting classifier*

Gradient Boosting Classifier (GBC) is a *supervised* learning algorithm that combines *weak learners* in an additive fashion to *boost* or improve the model creating a *strong learner*.[82,135] The basic procedure behind GBC involves beginning with a poorly performing model (or weak learner), let's say $M(x)$, which attempts to map features $x$ to the outputs $y$ from a set of training data, and does so with a considerable degree of error. The residual of $y$ and $M(x)$ is then used to train another weak learner, $G(x)$, which attempts to predict the error. The two models, $M(x)$ and $G(x)$, are then added together creating a slightly more accurate model, $H(x)$. This technique is then repeated, now using $H(x)$ as the original weak model. By repeating this process, the model can be improved incrementally until the desired accuracy is achieved.

When employing GBC, the training data takes the form of instances, i.e. features-label pairs $(x_i, y_i)$, $i = 1, 2, 3, \ldots, N$, where $x_i \in \mathbb{R}^n$, and $y \in \mathbb{R}$. The aim is to approximate the function $F^*(x)$—which maps the features $x$ to the corresponding label $y$—with $F(x)$, by minimizing the expected value of the loss function $L(y, F(x))$. Here, $F(x)$ is built additively by the following equation:

$$F_m(x) = F_{m-1}(x) + \rho_m h(x; a_m) \tag{4}$$

Where $\rho_m$ is the weight of the *base learner* function $h(x; a_m)$, with parameters $a = \{a_1, a_2, \ldots\}$, and $m = 1, 2, \ldots, M$, where $M$ is the number of iterations.[82,136] The base learner is a model of an ensemble of any type of weak learner, although it is common to use tree-based learners.[137] The starting model is initialized with the value

$$F_0(x) = \arg\min_{\rho} \sum_{i=1}^{N} L(y_i, \rho) \tag{5}$$

And the following iterations minimize

$$(\rho_m, a_m) = \arg\min_{\rho, h} \sum_{i=1}^{N} L(y_i, F_{m-1}(x_i) + \rho h(x_i; a)) \tag{6}$$

GBC solves eq. 6 in a two-step process. Firstly, $h(x; a)$ is determined using least squares

$$a_m = \arg\min_{a, \beta} \sum_{i=1}^{N} (r_{im} - \beta h(x_i; a))^2 \tag{7}$$

Where $r_{im}$ is the current *pseudo-residual*

$$r_{im} = -\left[\frac{\partial L(y_i, F(x_i))}{\partial F(x_i)}\right]_{F(x)=F_{m-1}(x)} \tag{8}$$

$h(x; a_m)$ has thus been trained by a new data set $(x_i, r_{im})$, which aims to predict the error of the previous iteration as mentioned above.

Lastly, $h(x; a_m)$ can be used to determine the optimal value of $\rho_m$

$$\rho_m = \arg\min_{\rho} \sum_{i=1}^{N} L(y_i, F_{m-1}(x_i) + \rho h(x_i; a_m)) \tag{9}$$

These results can then be inserted into eq. 4 yielding a model with arbitrary accuracy.[82,136–138]

Although the technique is quite robust, it can suffer from overfitting, without the proper modifications. A common modification is to add a *learning rate* parameter to eq. 4 giving

$$F_m(\bm{x}) = F_{m-1}(\bm{x}) + v\rho_m h(\bm{x}; \bm{a}_m) \tag{10}$$

Where the learning rate is confined to $0 \leq v \leq 1$. Although as $v$ gets smaller more iterations $M$ are required, in general the preferable configuration includes a large $M$ and $v < 0.1$. Overfitting can also be avoided (in situations where $h$ is a tree-based learner) by limiting the complexity of the trees by either limiting the maximum depth (generally limited to 3–5), the number of nodes, or the number of instances required to split a node. Another common method is to use random sub sampling without replacement (not allowing for the same instance to be chosen twice) when choosing the data with which to train $h$.[139,136]

### 4.3 Voting classifier

Voting Classifier (VC) is a machine learning technique that uses an ensemble of supervised learning algorithms/classifiers to determine the most accurate prediction by means of *voting*.[53,83,140] There are two main types of voting techniques: *hard/majority* voting and *soft* voting. In hard or majority voting, the decision is made based on the most frequent outcome amongst the ensemble of algorithms. For example, if there are three algorithms in the ensemble which are predicting a binary output of 1 or 0 and two out of the three output a 1, the VC would choose 1. In soft voting, the decision is made based on a weighted average of probabilities. Instead of the VC looking at a single output from each classifier in the ensemble, the probability of each classifier returning a certain output is considered. These probabilities are combined in a weighted sum to determine the final output of the VC. The weights given to each classifier are chosen based on the classifier's *importance* for the given output. For example, if there are once again three algorithms with outputs of either 1 or 0, the VC looks at the probability of each algorithm returning a 1 or a 0. For instance, algorithm one might have a 73% probability of returning a 1 and a 27% chance of returning a 0, and would be assigned a weight value for its importance in determining the specific result. These probabilities are then combined in a weighted sum with the values for the other two algorithms, and the output with the highest value is chosen (Fig. 3b).[141,142]

When utilizing VC, the training data must take the form of instances, i.e. features-label pairs $(\bm{x}_i, y_i)$, $i = 1, 2, 3, \ldots, N$, where $\bm{x}_i \in \mathbb{R}^n$, and $y \in \mathbb{R}$. For $M$ algorithms, $\bm{y}_i = \{y_{i1}, y_{i2}, \ldots, y_{iM}\}$ contains the output for each of the $M$ algorithms at each $i$-th instance. Depending on the type of VC being used, the outputs contained in $\bm{y}_i$ will either be used to find the most occurring output (hard voting), or the output with the highest weighted probability (soft voting). In the case of soft voting, the weight vector, $\bm{w}_i = \{w_{i1}, w_{i2}, \ldots, w_{iM}\}$, contains the weights for each of the $M$ algorithms at each $i$-th instance. Several approaches can be employed to determine the importance/weight of each algorithm.[83]

### 4.4 Neural networks

An Artificial Neural Network (ANN) or Neural Network (NN) for short, is a supervised learning method that is designed to resemble the structure of neurons and their connections to one another in the human brain. The network is made up of layers which contain multiple artificial neurons or nodes, each of which accepts an input and produces an output. The basic structure consists of an *input layer*, where the artificial neurons receive the inputted data, an *output layer*, which outputs the predictions, and a set of *hidden layers* between the input and output layers that are responsible for data processing (Fig. 3c).[93]

In a NN each node is connected to every node in the previous and following layer through an associated weight and activation threshold that are determined by training the network with known inputs-outputs sets. When training NNs the inputted data usually assumes the form of vectors or matrices; for example a matrix of pixels of an image, each of which is paired to an output or class, such as what the image depicts (cf. § 4.4.1). Once trained, the network returns the most likely output for a given set of inputs, on the basis of the determined activation thresholds and weights. Formally, the input for any given node is the weighted sum of all the outputs from the previous layer:

$$z_j = \sum_i w_{ij} x_i \qquad (11)$$

Where $w_{ij}$ is the weight for the path from the $i$-th neuron in the previous layer to the $j$-th neuron in the current layer, and $x_i$ is the output from the $i$-th neuron in the previous layer. At each stage, the $j$-th input $z_j$ is passed through an *activation function* before being outputted to the next layer of nodes. Simply put, the activation function determines whether the output for each node is activated or not (e.g. ON/OFF or 0/1) based on the received inputs, effectively establishing the contribution of every node of the network to the prediction process. Complex problems (e.g. classification or assignment problems) are therefore solved through relatively simple mathematical operations that render nodes active/inactive, for any specific set of inputs. There are several available activation functions; some of the most common are the rectified linear unit (ReLU), as well as the more conventional sigmoids (e.g. hyperbolic tangent) and the logistic function.[85] More complicated networks such as convolutional neural networks can have multiple and different types of hidden layers (cf. § 4.4.1).[143]

At the output layer, a *loss function* (e.g. the least-squares error function) is used to determine the error between the network's classification and the true classification. The gradient of the loss function with respect to the weights is then computed, and the results are used to update the values of the weights such that the overall loss is minimized. The process of updating the weights continues backwards through the network until the input layer is reached, this is known as *backpropagation*.

In the following subsections we briefly summarize two types of neural networks relevant to the demonstrations presented in this review.

*4.4.1 Convolutional Neural Networks*
A Convolutional Neural Network (CNN) is a supervised, deep learning technique, most commonly used for classifying image data, that utilizes layers of artificial neurons to analyze the inputted data with multiple levels of abstraction.[85,144] CNNs are a type of *Deep Learner*—a neural network consisting of many layers (Fig. 3c). The basic structure of a CNN matches that of a general NN, except for a few main characteristics. In CNNs the layers are organized in dimensions, they commonly include different types of layers—convolutional, pooling and non-linearity layers—and they are arranged such that not every node of each layer is connected to each node of the following layer. Convolutional neural networks take their name from the convolution (hidden) layers, which perform the convolution operation on two functions to produce a third one (see below).[143]

In general, CNNs are designed to interpret data in the form of arrays. For example, a simple black and white image might be represented by a 2D array of values, where each value represents the illuminance of a single pixel in the image. As the data is processed by each layer, it is passed to the neurons in the next layer (Fig. 3c). Each neuron's input in the layer is a weighted sum of the outputs from the neurons that are connected to it. Each connection has its own unique weight value which changes over time as the CNN learns. What each neuron does with the data depends on the type of layer that neuron belongs to.

In a convolutional layer, *filters* (or *kernels*) are applied to the data to extract key features from the image such as edges or shapes. Multiple layers with different filters can be used in the CNN to gain information about different aspects of the data. These filters are square matrices that are used to convolve the input matrix. For an input matrix of size $N \times N$, and a filter size of $F \times F$, the filter starts in the upper leftmost part of the input matrix, and the dot product between the filter and the $F \times F$ section of the input that it overlaps is taken and stored in the upper leftmost point in a new matrix. The filter then slides to the right by one unit, where the dot product is taken again for the new overlapped section and stored in the next entry of the new matrix. This process continues until the entire input matrix is traversed. The new resulting matrix will have dimensions $O \times O$, where[143]

$$O = 1 + (N - F)/S \tag{12}$$

Where $S$ is the *stride*, or the amount of units the filter moves by while traversing the input. In situations where many layers of convolution are used, it may be beneficial to employ *zero padding* so that the size of the matrix is not reduced to an unusable size. This technique simply involves adding one or more layers of zeros around the edge of the matrix to avoid reducing its size after the convolution. The input matrix gets scanned (convolved) with several different filters (kernels) each producing a corresponding *feature map* (effectively a new matrix) that get all combined as the output of the convolutional layer.

In a non-linearity layer, the input—a weighted sum of the previous output nodes that are connected—is passed through an *activation function* that determines which nodes are active and which are inactive (cf. § 4.4).

In pooling layers, the matrix size is reduced through a process known as *pooling*. In regards to images, this can be seen as a reduction in the resolution. There are a few different pooling operations, with the most prominent being *max-pooling*. Here, the input matrix is parsed using blocks of a chosen size (most commonly $2 \times 2$), where the maximum value in the block is recorded in a new matrix. This new matrix with a reduced size is then passed along as the output. The pooling operation effectively reduces the number of parameters and computation in the network, effectively shortening the training time, while also controlling overfitting.

CNNs learn through the aforementioned backpropagation process, where the network is traversed backwards (from the output layer to the input layer) in order to change the weights between neurons using, e.g., gradient descent to minimize a specified loss function. At any given layer the input is given by eq. 11. The output from any given layer is either the result of a convolution, a pooling operation, or the output from an activation function $f(z)$. At the output layer, the loss function $L(z, y)$ is calculated using the actual value in the (known) training set $y$ and the prediction $z$ made by the model. The gradient of the loss function is taken with respect to each of the weights used in the output layer. These weights are then adjusted such that the overall loss is minimized. This process continues backwards through the network until the input layer is reached.

*4.4.2 ADAptive LINear Element*
An Adaptive Linear Element (ADALINE) is a supervised, single-layer artificial neural network (ANN) for *binary* classification.[145] The basic structure of an ADALINE network follows that of a standard NN and consists of an input layer, where the neurons receive the inputted data, an output layer, where the model's results are presented, and a region of hidden layers in between where the predicted values are calculated and the learning/processing takes place (Fig. 3d).

When implementing ADALINE the training data is usually given in instances, i.e. features-label pairs $(x_i, y_i)$, $i = 1, 2, 3, ..., N$, where $x_i \in \mathbb{R}^n$, and $y \in \mathbb{R}$. The goal is to find a function $z$ that maps the features $x$ to the labels $y$ with the least error by changing the weights (Fig. 3d). For instance, for a single features-label pair,

$$z = \boldsymbol{w} \cdot \boldsymbol{x} + b \tag{13}$$

Where $b$ is the bias, $w$ is a vector containing the weights, and $x$ is the vector of features. The model learns by updating the weights $w$ through the following learning rule

$$W_{k,new} = w_{k,old} + \alpha(y-z)x_k \qquad (14)$$

Where $\alpha$ is a positive constant known as the *learning rate* (typically between 0 and 2), $z$ is the predicted value for the label (eq. 13), $y$ is the actual label, and $k$ denotes the component of the vector being updated (note that the bias is also updated using eq. 14 with $b$ in place of $w$). The weights are updated until, e.g., the mean squared error is minimized

$$MSE = \frac{1}{N}\sum_{i=1}^{N}(y_i - d_i)^2 \qquad (15)$$

Where the index $i$, once again, represents the $i$-th instance-label pair.

After multiple cycles of adjusting weights to minimize the error, the prediction $z$ from eq. 13 is computed one last time and passed through the threshold function, after which the classification is presented at the output.[146,53,147]

*4.5 Multi-feature linear regression*
Multi-feature linear regression or multiple linear regression (MLP) is a supervised learning technique that predicts the value of a label using multiple correlated features.[148,33] Here, the label or output is predicated by a linear, weighted sum of the features or inputs.

When using this technique the data consists of instances, i.e. features-label pairs $(x_i, y_i)$, $i = 1, 2, 3, \ldots, N$, where $x_i \in \mathbb{R}^n$, and $y \in \mathbb{R}$. The goal is to find the function $z(x)$ that most closely maps the features $x$ to the label $y$, while minimizing a certain cost function. For an individual predicted label

$$z = w_1 x_1 + w_2 x_2 + \cdots w_n x_n + b \qquad (16)$$

Where $b$ is the offset, $x_k$ is the $k$-th component of the $i$-th features vector $x_i$, and $w_k$ is the corresponding weight for said $k$-th component.

A standard cost function for MLP is the least squares error (LSE) cost function, or L2-norm loss function

$$LSE = \frac{1}{2N}\sum_{i=1}^{N}(y_i - z_i)^2 \qquad (17)$$

Where the subscript $i$ cycles through the different data sets (1 through $N$) used to train the model, and where the labels $y_i$ are the target values known from the training data. The weights $w_i$ are determined by minimizing eq. 17 through an optimization algorithm such as the *gradient descent* algorithm.[149]

*4.6 Bayesian machine learning*
Bayesian machine learning (BML) is a supervised, probabilistic approach that adheres to the framework of Bayesian statistics: *prior probability distributions* are updated into *posterior probability distributions* after processing new data.[90,89] In the context of machine learning, the Bayesian framework is adapted and invoked to either *i)* optimize a model from training data sets or *ii)* compare multiple models based on their respective marginal likelihood (i.e. the probability each model assigns to the observed data).

i) Model optimization
In this case, Bayesian statistics is used with the goal of optimizing the parameters of a chosen model, based on available data. Specifically, given a model $m$ of (unknown) parameters $\boldsymbol{\theta}$ and (observed) data $D$, Bayes theorem applied to machine learning is

$$P(\boldsymbol{\theta}|D,m) = \frac{P(D|\boldsymbol{\theta},m)P(\boldsymbol{\theta}|m)}{P(D|m)} \tag{18}$$

Where $P(D|\boldsymbol{\theta},m)$ is the likelihood of the parameters $\boldsymbol{\theta}$ in model $m$, while $P(\boldsymbol{\theta}|m)$ and $P(\boldsymbol{\theta}|D,m)$ are the prior and posterior probability, respectively, of the given data $D$.

In this framework, learning consists in using the data $D$ to update prior knowledge of the parameters $P(\boldsymbol{\theta}|m)$, into posterior knowledge of them, $P(\boldsymbol{\theta}|D,m)$. The posterior then becomes the prior for future data and the trained model can be used to predict unseen data $D_{new}$ through evaluating

$$P(D_{new}|D,m) = \int P(D_{new}|\boldsymbol{\theta},D,m)P(\boldsymbol{\theta}|m)d\boldsymbol{\theta} \tag{19}$$

*ii)* Models comparison

The Bayesian framework can also be applied to compare the effectiveness of different models. The idea is to identify a set $M$ of possible models for the data $D$ and assign prior belief $P(\boldsymbol{\theta}_j|m_j)$ to each model $m_j$ (each $j$-th model is parametrized by parameters $\boldsymbol{\theta}_j$). For the set of considered $M$ models, and associated prior beliefs in the appropriateness of each model $P(m_j)$, the posterior probability of each model $m_j$ is

$$P(m_j|D) = \frac{P(D|m_j)P(m_j)}{P(D)} \tag{20}$$

where

$$P(D) = \sum_{j=1}^{M} P(D|m_j)P(m_j) \tag{21}$$

And the marginal likelihood (i.e. the model evidence) of each model $m_j$ is

$$P(D|m_j) = \int P(D|\theta_j,m_j)P(\theta_j|m_j)d\theta_j \tag{22}$$

Where the integral is replaced by summation in discrete parameter spaces (note that the number of parameters $\dim(\boldsymbol{\theta}_j)$ does not need to be the same for each model). The different models can then be compared by comparing the respective probabilities given by (20). So, for instance, one can compare models $m_p$ and $m_q$ by simply calculating their relative probability, given the data $D$:

$$\frac{P(m_p|D)}{P(m_q|D)} = \frac{P(D|m_p)P(m_p)}{P(D|m_q)P(m_q)} \tag{23}$$

Note that $P(m_j|D)$ only refers to the probability relative to the set of $M$ models specified: it is not the absolute probability.

*4.7 Quantum Hamiltonian learning*

Quantum Hamiltonian learning (QHL) is a *supervised*, probabilistic technique that applies the method of Bayesian Learning to quantum systems.[125,91,150,59] Following from eq. 18,

$$P(\widehat{H}(x)|D) = \frac{P(D|\widehat{H}(x))P(\widehat{H}(x))}{P(D)} \tag{24}$$

where $D$ is the (experimental) data set, and $\widehat{H}(x)$ is a specific Hamiltonian with parameters $x$, i.e. $\widehat{H} = \widehat{H}(x)$. Equation 24 gives the probability that the current Hamiltonian $\widehat{H}(x)$ is the "true" Hamiltonian for the data set $D$. The process is repeated for each new set of acquired data, such that $P(\widehat{H}(x)) \leftarrow P(D|\widehat{H}(x))$; each repetition constitutes an update and the probability distribution should converge towards the true model parameters as the number of updates increases. This procedure works well for situations where the *likelihood function* $P(D|\widehat{H}(x))$ is easily determined and can be thought of as "classical Hamiltonian learning". However, for quantum systems it is often far too difficult to determine $P(D|\widehat{H}(x))$ with classical techniques, as these require time that scales polynomially in the Hilbert-space dimension and thus cannot simulate, practically, large quantum systems.[151,152] One possible alternative for determining the likelihood function is to perform quantum likelihood evaluation (QLE) or interactive quantum likelihood evaluation (IQLE) experiments.[150] The basic idea behind QLE/IQLE is to use a quantum simulator to conduct a multitude of simulations of the system using $\hat{H}(x)$ and collect the data from each. The quantum simulator can be a quantum computer or a special-purpose analog simulator, provided that its dynamics is sufficiently close to that of the ideal model and that the probability it computes would yield the observed measurement outcome. In QLE, the probability $P\left(D|\hat{H}(x)\right)$ is then the fraction of times $D$ occurs in a sufficiently-large number of simulations, which can make the process time-efficient given that to reach an error within $\varepsilon$, it suffices to draw $O(1/\varepsilon^2)$ samples for each probability.[91] In general, QHL can be an efficient strategy—for quantum simulation can be exponentially faster than alternative techniques—to tackle prohibitive problems in Hamiltonian estimation. This is especially true when combined with an accurate-enough knowledge of the physics of the system, which allows for a faster optimization of the QHL algorithm learning rate.

## 5. Conclusions and Outlook

Machine and quantum learning are convincingly establishing themselves as powerful approaches for tackling a wide variety of scenarios. For instance, they show remarkable aptness to solving specific types of problems such as those centered around pattern recognition, and/or data classification and clustering. In this review—through a selection of demonstrations—we have shown that diamond-based quantum measurements can effectively benefit from ML and QML strategies. Some of the most notable improvements include the speedup and increased accuracy of the measurements themselves. This is important as such improvements are not just desirable from a practical standpoint. Rather, they can lead to fundamental advantages such as, e.g., reaching resolutions or even capabilities otherwise only possible at cryogenic temperatures (cf. § 3.3). This, in turn, extends the scope and applicability of these quantum technologies. For completeness, we emphasize that while a subset of ML/QML offers a fundamentally different approach to traditional methods based on statistical inference (cf. §§ 2.3 and 3.5), in many cases the distinction is more conceptual (cf. § 2.1) as both ML/QML and inferential methods—ultimately—make use of statistical tools to respectively predict or determine measurement outcomes from available data. However, even in those cases where the distinction is mostly conceptual, we underscore that ML/QML can still be uniquely beneficial for certain quantum measurements, in particular for those where having access to complete sets of data might not always be possible or desirable (cf., e.g., §§ 3.1, 3.4).

As per potential future realizations beyond those discussed here, in general ML and QML strategies show potential for benefitting any diamond-based quantum technology that relies on the manipulation and readout of spin-optical defects—which is the basis of many quantum information, computing and metrology applications that use diamond as hardware material. This is a consequence of the demonstrated capacity of ML and QML algorithms to perform fast and accurate measurements even on sparse and incomplete datasets, again, common features in quantum measurements where short coherence and low signal-to-noise ratio might be scarce resources.

Other fields where ML and QML show great promise are material and device engineering and optimization. For these, learning-based strategies can contribute in two significant ways. They are an inexpensive way of, for instance, narrowing down the parameter space of specific variables involved in the fabrication process, and of testing—or even designing—target material/device properties that depend on such variables (cf. § 3.6). Additionally, ML and QML approaches are—intrinsically—automation-compatible, which is desirable for large-scale and fast material/device fabrication, and, in general, for any task requiring the repeated or programmable execution of measurements and/or operations.

**Acknowledgements**

The Natural Sciences and Engineering Research Council of Canada (via RGPIN-2021-03059, DGECR-2021-00234, USRA-574869-2022) and the Canada Foundation for Innovation (via John R. Evans Leaders Fund #41173) are gratefully acknowledged.